%% file: main.tex
\documentclass[a4paper,twoside]{article}

\usepackage{epsfig}
\usepackage{subfigure}
\usepackage{calc}
\usepackage{amssymb}
\usepackage{amstext}
\usepackage{amsmath}
\usepackage{amsthm}
\usepackage{multicol}
\usepackage{pslatex}
\usepackage{apalike}
\usepackage{SciTePress}
\usepackage{url}
\usepackage[small]{caption}
\usepackage[latin1]{inputenc} 
\usepackage{focus} 

\newcommand{\focus}{\Focus}

\begin{document}

\newcommand{\msfn}[1]{\textsf{\small#1}}
\newcommand{\msf}[1]{\textsf{\footnotesize#1}}
\newcommand{\mbsf}[1]{\textsf{\textbf{\footnotesize#1}}}
\newcommand{\autofocusString}{\textsc{AutoFocus}}
\newcommand{\autofocus}{\autofocusString\xspace} 
\newcommand{\autofocusIII}{\autofocus~3\xspace} 
\newcommand{\aft}{\autofocusIII}

\title{Do we really need to write documentation for a system?  \subtitle{
CASE tool add-ons: generator+editor for a precise documentation
} }

\author{\authorname{Maria Spichkova\sup{1}, Xiuna Zhu\sup{1} and Dongyue Mou\sup{2}}
\affiliation{\sup{1}Institut für Informatik, Technische Universität München, Boltzmannstr. 3, 85748 Garching bei München, Germany}
\affiliation{\sup{2}Fortiss GmbH, Guerickestr. 25, 81735 München, Germany}
\email{\{spichkov, zhux\}@in.tum.de, mou@fortiss.org}
}


\abstract{
One of the common problems of system development projects is that the
system documentation is often outdated and does not describe the latest version of the system.
The situation is even more complicated if we are speaking not about a natural language description of the system, but about its formal specification.
In this paper we discuss how the problem could be solved by updating the documentation automatically, by generating a new formal specification from the model if the model is frequently changed. 
}

\onecolumn \maketitle \normalsize \vfill

\section{\uppercase{Introduction}}
\label{sec:introduction}

System documentation is  claimed to be an important part of the development process but is very often considered by industry as a secondary appendage to the main part of the development -- modeling and implementation.  
The implementation of the system is what we would like to get at the end, a model of the system is what helps us to simulate and to implement the system, but the documentation is mostly important   ``only" for the cooperation of experts of different disciplines, for maintenance, and for reuse of the system. 
This causes the situation that the system documentation is often  outdated and does not describe the latest version of the system: system requirements documents and the general systems description are not updated according to the system's or model's modifications -- sometimes because this update is overlooked, sometimes on purpose, because of  timing or costs constraints on the project.

This problem is even more complicated if we are speaking not about a natural language description of the system, but about its formal specification. 
On the one hand, a formal specification provides a system description that is 
much more precise than the natural language one and it can help to solve a lot of specification/documentation problems, 
but on the other hand, it takes much more time to write a formal specification than an informal natural language description. 
Unfortunately, we should acknowledge that dealing with formal methods often assumes that only two factors must be satisfied: the method must be sound and give such a representation, which is short and beautiful from the mathematical point of view, without taking into account any question of readability, usability, or tool support. This leads to the fact that formal methods are treated by most engineers as ``something that is theoretically important but practically too hard to understand and to use", where even small changes of a formal method can make it more understandable and usable for an average engineer. 
Our approach Human Factors of Formal Methods, $HF^2M$~\cite{hffm_spichkova} focuses on human factors in formal methods used within the  specification phase of a system development process \cite{dentum_tb2,dentum_tb}: during  requirements specification and during the development of a system architecture. 

However, even a readable formal  specification is hard to keep up to date if the system model is frequently changing during the modeling phase of the development. This problem could be solved by appropriate automatization: if we make these updates automatically or   generate new (updated) formal specifications from the model. 
To allow a simple and intuitional design of distributed systems and applications,  Computer-Aided Software Engineering (CASE) 
tools are widely used. The CASE tools could also help to solve the problem with the system documentation: 
in this paper we present an extension of  the AutoFocus CASE Tool by add-ons that allow to generate the formal specification according to the ideas presented in~\cite{hffm_spichkova} as well as to edit a generated formal specification or write a specification using the predefined templates.

\section{\uppercase{CASE tool Model: AutoFOCUS}}
\input{af}

\section{\uppercase{Formal Specification: FOCUS}}
\input{focus_optimized}

\input{generators}

\section{\uppercase{Conclusions and Future work}}
\label{sec:conclusion}

In this paper we introduce a user-friendly tool support for a formal system documentation.
Firstly, we discuss how the problem with outdated system documentation could be solved by making the documentation updates automatically, by generating a new formal specification from the model if the model is frequently changed. After that 
an extension of the AutoFocus CASE Tool is presented: the add-ons that allow 
\begin{itemize}
\item
to generate a formal \Focus specification by taking into account the theories of  human factors, 
\item
 to edit a generated formal specification, and
 \item
  write a specification using the predefined templates.
\end{itemize}
The presented results can be integrated into the development methodology for verified software systems~\cite{VerisoftXT_FMDS,af3paperSE}.
Using this approach, one can go further and verify properties of a system  in a formal way according to the methodology ``\Focus on Isabelle''\cite{spichkova}, 
by translating the \Focus specifications to the semiautomatic theorem prover Isabelle/HOL~\cite{npw}, an interactive semi-automatic theorem prover,  and using the Isabelle tool to make the proofs. Using an AutoFocus model one can also take an advantage of the user-friendly verification environment for model cheking~\cite{Campetelli11}.

\bibliographystyle{apalike}
{\small{

}}

\vfill
\end{document}

%% file: af.tex
\label{sec:af}

\aft  \cite{af3paper,IntegratingFDT:FM1999,Schaetz:Kluwer2004} is a scientific CASE tool prototype\footnote{\url{http://af3.fortiss.org}}  
implemented on top of the Eclipse\footnote{\url{http://www.eclipse.org}}  platform, and
based on a graphical notation and a restricted version of the formal semantics of the \focus specification and modeling language, presented in Section~\ref{sec:focus},  in particular the
time-synchronous frame. 

The system structure specification captures in \aft the static aspects of the system description: 
a network of communicating components working in parallel.
Each component has a syntactic interface described by a set of ports, and the network of components is formed by connecting ports with channels. 
Each port is either an input or an output port, has a data type
and an initial value. 
Furthermore, each component is declared to be weakly causal or strongly causal: 
weak causality models instantaneous reaction, while strong causality models a 
reaction with some delay.

System structure specifications  in \aft may be separated into hierarchic views in order to deal with larger models. 
Components can be refined into a set of sub-components introducing both local communication
and communication to the environment through the interface of the parent component. 
Atomic components have their behavior specified using one of the following variants: a stateful
\emph{automaton specification} or a stateless \emph{function specification}.

Specifying a system in \aft, we obtain an executable model, which can be validated
using the \aft simulator to get a first impression of the system under development
and possibly find implementation errors that we introduced during the manual 
transformation
of the requirements into a \aft model. Automatisation of this transformation is a future work.

%% file: focus_optimized.tex
\label{sec:focus}

The formal background on {\Focus}\footnote{\url{http://focus.in.tum.de}} and its extensions are presented in \cite{focus} and \cite{spichkova}. Here
we use an optimized version of \Focus developed according to the
 $HF^2M$ approach presented in ~\cite{hffm_spichkova}:
it allows us to have shorter and more readable formal specifications. 
In many cases even not very complicated optimization changes of a specification method can make it more understandable and usable. Such a simple kind of optimization is often overlooked just because of its obviousness, and it would be wrong to ignore the possibility to optimize the language without much effort. For example, simply adding an enumeration to the formulas in a large formal specification makes its validation on the level of specification and discussion with co-operating experts much easier. 
The first results of visual optimization of Focus specifications are presented in~\cite{spichkova_processes}. 
 
A specification scheme of \Focus is inspired by specification approaches like Z (see \cite{Spivey_88}), 
but the \Focus framework is much more powerful and expressive:  
it supports a variety of specification styles which describe system
components by logical formulas or by diagrams and tables representing logical formulas. 
\Focus  
has an integrated notion of time and modeling techniques for unbounded networks,  
provides a number of specification techniques for distributed systems and concepts of refinement. 
For example, the B-method~\cite{bbook} is used in many publications 
on fault-tolerant systems, but 
it has neither graphical representations nor integrated notion of time. 
Moreover, the B-method also is slightly more low-level and more focused on the refinement to code rather than formal specification.  
Formal specifications of real-life systems can become very large and complex, 
and are as a result hard to read and to understand. 
Therefore, it is too complicated to start the specification process in some low-level framework, First-Order or Higher-Order Logic  directly: 
the graphical specification style is essential here.

The central concept in \Focus is a \emph{stream} representing a 
communication history of  a {\it directed channel} between components. 
A system in \Focus is represented by its components that are 
connected by channels, 
and are described in terms of its input/output
behavior. Thus, the components can interact and also work independently of each other.
The channels in this specification framework are \emph{asynchronous communication links} 
without delays. They are \emph{directed} and generally assumed to be \emph{reliable},
 and \emph{order preserving}. Via these channels components
 exchange information in terms of \emph{messages} of specified types.

A specification can be elementary or composite -- composite specifications are
built hierarchically from the elementary ones. 
Any specification characterizes the relation between the
\emph{communication histories} for the external \emph{input} and \emph{output channels}: 
the formal meaning of a specification is exactly the \emph{input/output relation}.

%% file: generators.tex
\begin{figure*}[ht!]
\centering \includegraphics[scale=0.6]{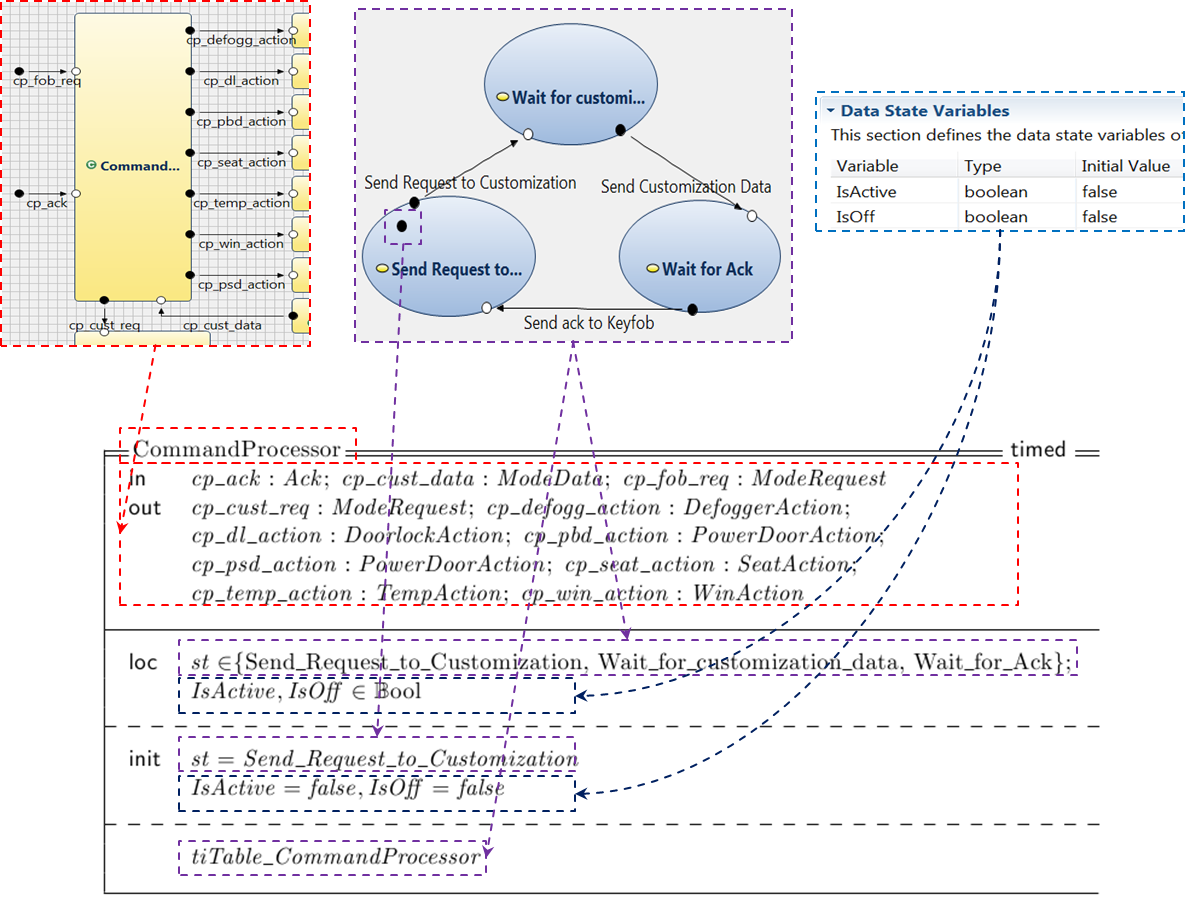}
  \caption{Generation of the component specification}
  \label{fig:component}
  ~\\~
\end{figure*}

\begin{figure*}[ht!]
\centering \includegraphics[scale=0.35]{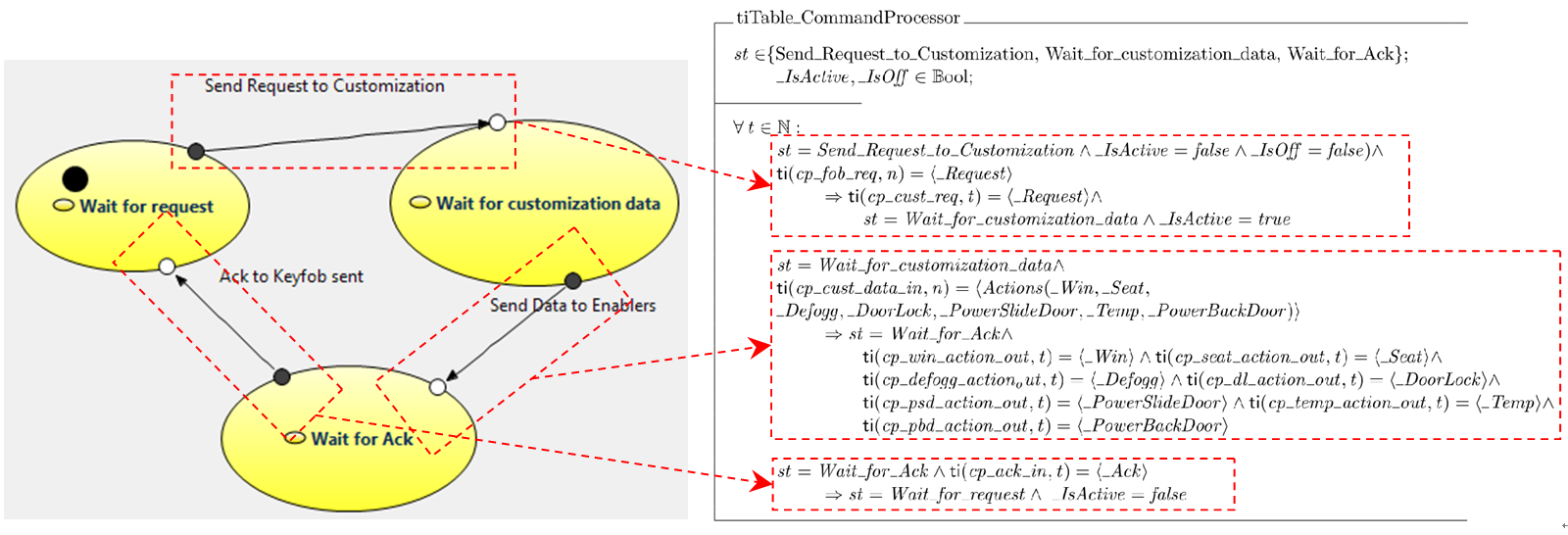}
  \caption{Generation of the \Focus timed state diagram (plain text) specification}
  \label{fig:automaton}
  ~\\~\\~
\end{figure*} 

\section{\uppercase{Documentation artifacts}}
\label{sec:generators}

The \Focus generator produces a specification of the model by representing the formal specification in LaTeX  according to the predefined templates. 
We choose for the specification generation the  most general style of a \Focus specification, that  is, an Assumption/Guarantee style: 
where a component is specified in terms of an assumption and a guarantee: whenever input from the environment behaves in accordance
with the assumption $\msfn{asm}$, the specified component is required to fulfill the
guarantee $\msfn{gar}$.

\begin{figure*}[ht!]
\centering \includegraphics[scale=0.7]{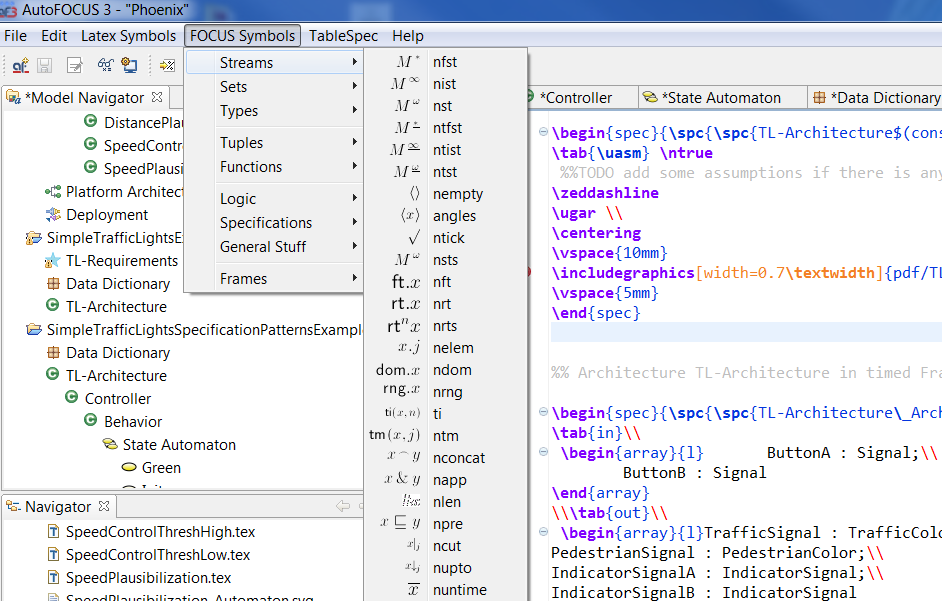}
  \caption{\Focus editor: Part of a specification template}
  \label{fig:FocusEditor}
  ~\\~
\end{figure*}

For illustration,  Figure~\ref{fig:component} shows how a formal specification of an elementary component is generated from the model:
\begin{itemize}
\item 
The interface part of the \Focus specification is generated from the  external syntactic interface of the AutoFocus component (red dotted block).
\item
The \emph{init}-part of the specification contains all the information about the initial values of all the local variables for the component state and data. 
The initial state is marked in the model by the black dot --  name of this state provides the initial value of the $st$ varible, where the initial values of the data state variables are represented separately (blue dotted block).
\item
The main part of the specification is generated from the AutoFocus State Transition Diagram (lila dotted block).
Firstly, the set of component states builds the corresponding data type of the state variable $st$; 
Secondly,  the state automaton corresponds to the guarantee-part of the specification and can be represented by the \Focus timed state diagram (represented by a \Focus automaton as well as by a plain text specification as shown on Figure~ \ref{fig:automaton}) or by a timed table~(see also \cite{spichkova_avocs_2012}). Each transition of the automaton is  described by a formula (plain text variant) or by a line of the timed table.
\end{itemize}
The \Focus specification of a composite component can be represented in two ways: 
\begin{itemize}
\item 
graphically, almost similar to the AutoFocus representation, or 
\item by plain text:  
the  guarantee-part of the specification describes $(i)$ which local channels the composite component has and $(ii)$ how the subcomponents are connected via these channels.
\end{itemize}
~\\
The generated specification (.tex files) can also be optimized or extended manually, using the \Focus editor we have implemented and consequently transformed to  the PDF- or DVI-document. Figure~\ref{fig:FocusEditor}  shows a part of a specification template. 
This editor inherits most of the functions from the  open source plugin TeXlipse\footnote{\url{http://texlipse.sourceforge.net}} (e.g., the syntax check of the specification as well as syntax highlighting, code folding, etc.), and is extended by additional features such as 
\begin{itemize}
\item
 \Focus operators as well as  the main \Focus frames: component and function specification, 
 \item
 \Focus timed tables, 
 \item
 predefined data types and streams,
 \item
 tool box for the predefined \Focus operators, which allows a quick access to the most important features of the formal language.
 \end{itemize} 
Thus, this add-on provides a user-friendly interface which, on the one hand, is oriented on the features of the \Focus language, and on the other hand, 
does not require any special sophisticated knowledge.

Both add-ons, the formal specification generator and the editor,  are planed to be a part of the general distibution of  \aft.

%% file: main.bbl
\begin{thebibliography}{}

\bibitem[Abrial, 1996]{bbook}
Abrial, J.-R. (1996).
\newblock {\em {The B-book: assigning programs to meanings}}.
\newblock Camb.\,\!Univ.\,\!Press.

\bibitem[Broy and St{\o}len, 2001]{focus}
Broy, M. and St{\o}len, K. (2001).
\newblock {\em Specification and Development of Interactive Systems: Focus on
  Streams, Interfaces, and Refinement}.
\newblock Springer.

\bibitem[Campetelli et~al., 2011]{Campetelli11}
Campetelli, A., H{\"o}lzl, F., and Neubeck, P. (2011).
\newblock User-friendly model checking integration in model-based development.
\newblock In {\em 24th International Conference on Computer
  Applications in Industry and Engineering}. The International Society for Computers and Their Applications.

\bibitem[Feilkas et~al., 2009]{dentum_tb}
Feilkas, M., Fleischmann, A., H\"olzl, F., Pfaller, C., Rittmann, S.,
  Scheidemann, K., Spichkova, M., and Trachtenherz, D. (2009).
\newblock {A Top-Down Methodology for the Development of Automotive Software}.
\newblock Technical Report TUM-I0902, {TU M{\"u}nchen}.

\bibitem[Feilkas et~al., 2011]{dentum_tb2}
Feilkas, M., Hölzl, F., Pfaller, C., Rittmann, S., Schätz, B., Schwitzer, W.,
  Sitou, W., Spichkova, M., and Trachtenherz, D. (2011).
\newblock {A Refined Top-Down Methodology for the Development of Automotive
  Software Systems - The KeylessEntry-System Case Study}.
\newblock Technical Report {TUM-I1103}, TU M\"unchen.

\bibitem[H\"{o}lzl and Feilkas, 2010]{af3paper}
H\"{o}lzl, F. and Feilkas, M. (2010).
\newblock AutoFocus 3: A scientific tool prototype for model-based development
  of component-based, reactive, distributed systems.
\newblock In {\em Proc. of the 2007 International Dagstuhl conference on
  Model-based engineering of embedded real-time systems}, pp.
  317--322.

\bibitem[Nipkow et~al., 2002]{npw}
Nipkow, T., Paulson, L.~C., and Wenzel, M. (2002).
\newblock {\em {Isabelle/HOL -- A Proof Assistant for Higher-Order Logic}},
  volume 2283 of {\em LNCS}.
\newblock Springer.

\bibitem[Sch\"atz, 2004]{Schaetz:Kluwer2004}
Sch\"atz, B. (2004).
\newblock {Mastering the Complexity of Reactive Systems: the \textsc{AutoFocus}
  Approach}.
\newblock In Kordon, F. and Lemoine, M., editors, {\em {Formal Methods for
  Embedded Distributed Systems: How to Master the Complexity}}, pages 215--258.
  {Kluwer Academic Publishers}.

\bibitem[Sch{\"a}tz and Huber, 1999]{IntegratingFDT:FM1999}
Sch{\"a}tz, B. and Huber, F. (1999).
\newblock {Integrating Formal Description Techniques}.
\newblock In Wing, J.~M., Woodcock, J., and Davies, J., editors, {\em FM'99},
  volume 1709 of {\em {LNCS}}, pages 1206--1225. Springer.

\bibitem[Spichkova, 2007]{spichkova}
Spichkova, M. (2007).
\newblock {\em {Specification and Seamless Verification of Embedded Real-Time
  Systems: FOCUS on Isabelle}}.
\newblock PhD thesis, {TU M{\"u}nchen}.

\bibitem[Spichkova, 2011]{spichkova_processes}
Spichkova, M. (2011).
\newblock {Focus on processes}.
\newblock Tech. Report TUM-I1115, {TU M{\"u}nchen}.

\bibitem[Spichkova, 2012a]{spichkova_avocs_2012}
Spichkova, M. (2012a).
\newblock {Focus on Time}.
\newblock In {\em {Proceedings of the 12th International Workshop on
  Automated Verification of Critical Systems (AVoCS 2012)}}.

\bibitem[Spichkova, 2012b]{hffm_spichkova}
Spichkova, M. (2012b).
\newblock {Human Factors of Formal Methods}.
\newblock In {\em {IADIS Interfaces and Human Computer Interaction 2012
  (IHCI 2012)}}.

\bibitem[Spichkova et~al., 2012]{VerisoftXT_FMDS}
Spichkova, M., H\"olzl, F., and Trachtenherz, D. (2012).
\newblock {Verified System Development with the AutoFocus Tool Chain}.
\newblock In {\em 2nd Workshop on Formal Methods in the Development of
  Software}, WS-FMDS.

\bibitem[Spivey, 1988]{Spivey_88}
Spivey, M. (1988).
\newblock {Understanding Z -- A Specification Language and Its Formal
  Semantics}.
\newblock Cambridge Tracts in Theoretical Computer Science 3. Camb. Univ.
  Press.

\bibitem[Thyssen et~al., 2010]{af3paperSE}
Thyssen, J., Ratiu, D., Schwitzer, W., Harhurin, A., Feilkas, M., and Thaden,
  E. (2010).
\newblock {A System for Seamless Abstraction Layers for Model-based Development
  of Embedded Software}.
\newblock In {\em Software Engineering}, SE'10.

\end{thebibliography}
